\documentclass[useAMS,usenatbib,usegraphicx]{mn2e}
\usepackage{amsmath}
\usepackage{amssymb}
\usepackage{multirow}

\def\simlt{\lower.5ex\hbox{$\; \buildrel < \over \sim \;$}}
\def\simgt{\lower.5ex\hbox{$\; \buildrel > \over \sim \;$}}

\newcommand{\bd}{\begin{displaymath}}
\newcommand{\ed}{\end{displaymath}}
\newcommand{\be}{\begin{equation}}
\newcommand{\ee}{\end{equation}}
\newcommand{\beqa}{\begin{eqnarray}}
\newcommand{\eeqa}{\end{eqnarray}}

\newcommand{\vbc}{v_{\rm bc}}

\newcommand{\cool}{{\rm cool}}

\title[Complete 21-cm history from the first stars] {Complete
  history of the observable 21-cm signal from the first stars during the pre-reionization
  era} \author[Fialkov, Barkana,
Pinhas \& Visbal] {Anastasia Fialkov$^{1}$\thanks{E-mail:
    anastasia.fialkov@gmail.com},
  Rennan Barkana$^{1}$, Arazi Pinhas$^{2}$, Eli Visbal$^{3, 4}$ \\
  $^{1}$ Raymond and Beverly Sackler School of Physics and Astronomy,
  Tel Aviv University, Tel Aviv 69978, Israel \\ $^{2}$ Department of
  Physics and Astronomy, University of Pennsylvania, 209 South 33rd
  Street, Philadelphia, PA 19104-6396, USA \\ $^{3}$ Jefferson
  Laboratory of Physics, Harvard University, Cambridge, MA 02138, USA
  \\ $^{4}$ Institute for Theory $\&$ Computation, Harvard University,
  60 Garden Street, Cambridge, MA 02138, USA }

\begin{document}
\pagerange{\pageref{firstpage}--\pageref{lastpage}} \pubyear{2013}
\maketitle

\label{firstpage}

\begin{abstract}
  We present the first complete calculation of the history of the
  inhomogeneous 21-cm signal from neutral hydrogen during the era of
  the first stars. We use hybrid computational methods to capture the
  large-scale distribution of the first stars, whose radiation couples
  to the neutral hydrogen emission, and to evaluate the 21-cm signal
  from $z \sim 15-35$. In our realistic picture large-scale
  fluctuations in the 21-cm signal are sourced by the inhomogeneous
  density field and by the Ly$\alpha$ and X-ray radiative backgrounds.
  The star formation is suppressed by two spatially varying effects:
  negative feedback provided by the Lyman-Werner radiative background,
  and supersonic relative velocities between the gas and dark matter.
  Our conclusions are quite promising: we find that the fluctuations
  imprinted by the inhomogeneous Ly$\alpha$ background in the 21-cm
  signal at $z \sim 25$ should be detectable with the Square Kilometer
  Array.
\end{abstract}


\section{Introduction}

The history of the Universe between hydrogen recombination and the end
of the epoch of reionization remains mostly unobserved
\citep{Barkana:2001}. Luckily, this era can be probed by measuring the
redshifted 21-cm emission of neutral hydrogen. The state of the art
radio experiments (e.g., the Murchison Widefield Array (MWA)
\citep{MWAref} and the LOw Frequency ARray (LOFAR) \citep{LOFARref})
are designed to constrain the 21-cm signal from the epoch of
reionization, probing the redshift range $z\sim 7-15$; whereas next
generation experiments, such as the Square Kilometer Array (SKA)
\citep{Carilli:2004}, will probe the epoch of primordial star
formation out to $z \sim 30$.

Prior to reionization, radiation emitted by the first galaxies drove a
chain of significant cosmic transitions, which are important to test
observationally.  In particular, the X-ray background heated the gas
above the Cosmic Microwave Background (CMB) temperature; Ly$\alpha$
photons coupled the 21-cm spin temperature to the gas kinetic
temperature through the Wouthuysen-Field effect
\citep{Wouthuysen:1952, Field:1958}, thus making the epoch of
primordial star formation available for observations; and, finally,
photons in the Lyman-Werner (LW) band dissociated molecular hydrogen.
The relative importance of the three radiative backgrounds is studied
in this Letter.

The first stars are believed to form via the cooling of molecular
hydrogen (the lowest temperature coolant in the metal-free early
Universe), in halos of mass above the threshold minimum cooling mass
$M_{\cool} \sim 10^5M_\odot$ \citep{Tegmark:1997}. It is possible,
however, that H$_2$ cooling leads to very inefficient star formation,
in which case stars would only form via cooling of atomic hydrogen in
more massive halos of $M_{\cool} \sim 10^7M_\odot$.  According to the
common picture of hierarchical structure formation, in which small
objects form first, all these halos are expected to be rare (and
highly clustered) at high redshift, as they can form only in high
density peaks of matter fluctuations \citep{Barkana:2004}. The rarity
is even more extreme due to two effects, which are usually omitted but
which we include here. First is the effect of the LW background, which
destroys molecular hydrogen and increases the minimum cooling mass,
thus acting as negative feedback and delaying star formation.
Unfortunately, modeling of this effect in simulations is rather
limited at the moment. In fact, the effect of the feedback on
$M_{\cool}$ has been studied only in an unrealistic setup in which the
LW intensity during the entire process of gravitational collapse of a
halo is fixed \citep{Machacek:2001,Wise:2007,O'Shea:2008}.  Since in
reality the LW intensity rises exponentially fast with time
\citep{Holzbauer:2011}, there is some uncertainty in how to normalize
to the simulation results.  In \citet{Fialkov:2013} we dealt with this
by parameterizing the strength of the feedback and comparing the
effect of totally negligible, inefficient (``weak'') and highly
efficient (``strong'') LW feedbacks, as well as the case of completely
saturated feedback (i.e., no H$_2$ cooling, when stars are formed only
via cooling of atomic hydrogen).

The second important process that we include is the effect of
high-redshift inhomogeneous supersonic relative velocities between the
baryons and the cold dark matter, denoted by $\vbc$, on primordial
star formation, which only recently was noticed by
\citet{Tseliakhovich:2010}. The authors of this paper showed that
relative velocities were coherent on scales of several Mpc and their
power spectrum exhibited strong Baryon Acoustic Oscillations (BAOs;
see also \citet{Dalal:2010}).  The effect of $\vbc$
on star formation at high redshifts can be dramatic: the coherent
stream of baryons through a dark matter halo hinders accretion and
redistributes the gas density within the halo.  As a result, a heavier
halo is needed to form a dense cloud of gas that will allow star
formation \citep{Greif:2011,Stacy:2011}. Thus, in regions where $\vbc$
is high $M_{\cool}$ is boosted, which delays star formation and
increases galaxy clustering \citep{Tseliakhovich:2011,Fialkov:2011}.

Unfortunately, the strong fluctuations on large scales in the number
of stars, modified by the LW and $\vbc$ effects \citep{Fialkov:2013},
make it difficult to predict the 21-cm signal from the epoch of
primordial star formation. \citet{Visbal:2012} and
\citet{Fialkov:2013} used a hybrid numerical method (for a detailed
discussion of the structure of our simulation see section S1 of the
Supplementary Information of the former paper, and references within)
to show that the signal from $z \sim 20$ exhibits strong heating
fluctuations on $\sim 100$ Mpc scales, which can be detected using
present day observational technology.

In this Letter we extend the work of \citet{Fialkov:2013} by adding a
self-consistent inhomogeneous Ly$\alpha$ background, which allows us
to compile a more complete history of the 21-cm signal from high
redshifts ($10\lesssim z \lesssim 40$). We use the hybrid methods to
evolve in the same framework the large-scale distribution of stars and
the three radiative backgrounds emitted by these stars (including
Ly$\alpha$, LW and X-rays and neglecting the ionizing radiation which
becomes important at lower redshifts), and to estimate the 21-cm
signal from $10\lesssim z \lesssim 40$.  We use the standard set of
cosmological parameters \citep{WMAP7}, a star formation efficiency of
$f_* = 10\%$ (with additional $\log(M)$ suppression at small masses
\citep{Machacek:2001}), LW parameters as explained in detail in
\citet{Fialkov:2013}, and Ly$\alpha$ parameters as in
\citet{Barkana:2005}. We assume that the X-ray luminosity is
characterized by a power law with a spectral index $\alpha = 1.5$ and
an X-ray photon efficiency of $\xi_X = 10^{57}M_{\odot}^{-1}$ as in
\citet{21cmfast}, based on observed starbursts at low redshift.  In
addition, we account approximately for the redistribution of the UV
photons due to their scattering in the wings of the Ly$\alpha$ line of
hydrogen \citep{Naoz:2008} and for the $\vbc$-dependent suppression of
the filtering mass \citep{Naoz:2013}.

\section{Global History}

\begin{figure}
\includegraphics[width=3.4in]{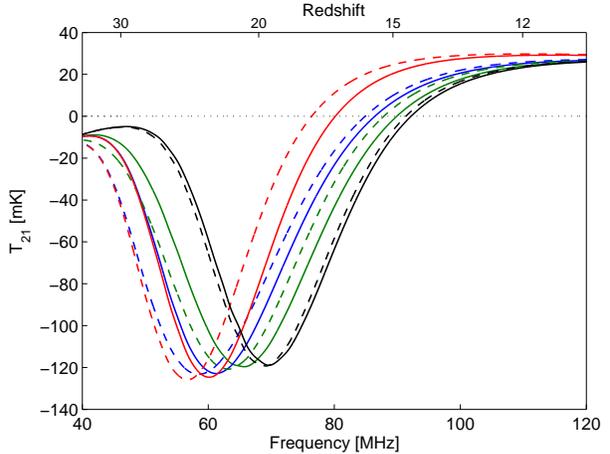}
\caption{Global spectrum of the redshifted 21-cm brightness
  temperature with (solid) and without (dashed) $\vbc$, for the cases
  of no H$_2$ cooling (black), strong (green) and weak (blue) LW
  feedback and pure H$_2$ cooling without the negative feedback (red).
  Note that the red dashed line is the previously standard case for
  small halos.}
\label{fig:T21}
\end{figure}

The global spectrum of the redshifted 21-cm signal, and in particular
the deep absorption feature expected to be observed at $\sim 70$
MHz
is a main target of some of the next generation experiments, such as
the Large-aperture Experiment to detect the Dark Ages (LEDA)
\citep{LEDA}. The exact shape and the central frequency of this
feature both depend on the details of the evolution history of the
Universe and in particular on the intensity and relative timing of the
X-ray, LW and Ly$\alpha$ radiative backgrounds emitted by the first
stars (see Figure~\ref{fig:T21} and Table~\ref{Tab:Global}).

\begin{table*}
\begin{center}
\begin{tabular}{ |l |r | r | r | r | r | r |}
\hline
Feedback, $\vbc$ & $\frac{dT_{21}}{d\nu}_{\rm{max,1}}~ \left[\frac{\textrm{mK}}{\textrm{MHz}}\right]$ &
$\nu_{\rm{max,1}}$ [MHz] (z)& $T_{21,\rm{min}}$ [mK]& $\nu_{\rm{min}}$ [MHz] (z)& $\frac{dT_{21}}{d\nu}_{\rm{max,2}}~ \left[\frac{\textrm{mK}}{\textrm{MHz}}\right]$ &  $\nu_{\rm{max,2}}$ [MHz] (z)\\
\hline
None, no $\vbc$ & -11.6 & 48.5 (28.4) & -125.9 & 56.7 (24.2) & 9.0 & 64.6 (21.1)\\
None, $\vbc$ & -11.3& 53.0 (26.0) & -124.7 & 60.1 (22.8) & 8.7 & 68.2 (20.0) \\
Weak, no $\vbc$ & -10.3    &  48.4 (28.5) &  -123.2 & 58.5 (23.4) & 6.5 & 69.9 (19.4) \\
Weak, $\vbc$ & -10.5 &  53.0 (26.0) &  -122.9 & 61.2  (22.3) & 6.9 & 70.9 (19.2) \\
Strong, no $\vbc$ & -8.6   & 53.1 (25.9)    & -120.8 & 63.2 (21.6)  & 6.9 & 73.7 (18.4) \\
Strong, $\vbc$ & -8.6   & 55.9 (24.5)    & -119.6 & 65.7 (20.7) & 6.9 & 74.7 (18.1) \\
No H$_2$, no $\vbc$& -9.8& 60.3 (22.7) & -119.3 & 69.2 (19.7) & 7.6 & 78.3 (17.3)\\
No H$_2$ & -9.7& 62.8 (21.7) & -119.0 & 69.6 (19.5)& 7.5 & 78.6 (17.2)\\
\hline
\end{tabular}
\caption{\label{Tab:Global}Characteristics of the absorption trough in
  the global spectrum of the redshifted 21-cm signal, shown in
  Figure~\ref{fig:T21}. Columns from left to right: the steepest
  negative slope of the spectrum; corresponding frequency (redshift);
  the minimum value of the brightness temperature; corresponding
  frequency (redshift); the steepest positive slope; corresponding
  frequency (redshift). }
\end{center}
\end{table*}

Both the LW feedback \citep{Fialkov:2013} and relative velocities
\citep{Fialkov:2011} delay star formation, which affects all the
milestones of the global evolution of the Universe. For instance, the
redshift of the heating transition, i.e., the moment when the gas
heats up to the temperature of the CMB, is delayed by up to $17\%$ in
$1+z$ when $\vbc$ and the feedback are applied \citep{Fialkov:2013}.
To highlight the effect of the feedback on the important milestones of
the high-redshift evolution history of the Universe (accounting also
for $\vbc$), we list here the redshifts of the Ly$\alpha$ transition
$z_{Ly\alpha}$ (defined as when the coupling of the spin temperature
to the gas temperature becomes significant, specifically when the mean
Ly$\alpha$ intensity equals the thermalization rate from
\citet{Madau:1997}), the heating transition $z_h$ (when the mean gas
temperature equals that of the CMB), and the epoch when LW feedback
becomes significant $z_{LW}$ (which is when the LW intensity equals
$J_{21} = 0.1$ in units of $10^{-21}$ erg s$^{-1}$ cm$^{-2}$ Hz$^{-1}$
sr$^{-1}$). Our computational method allows us to study the relative
timing between these transitions, which is considered highly uncertain
in the literature \citep{Pritchard:2012}.  We find that (for our
choice of model parameters) first to happen is the Ly$\alpha$
transition. The characteristic redshifts are $1+z_{Ly\alpha}= 26.8$ in
the case of H$_2$ cooling without the negative feedback,
$1+z_{Ly\alpha}= 26.5$ and 24.6 in the case of weak and strong LW
feedback respectively, and $1+z_{Ly\alpha}= 22.4$ in the case of no
H$_2$ cooling.  The LW feedback becomes important at intermediate
redshifts of $1+z_{LW}= 23.6$ for the strong feedback and $1+z_{LW}=
19.2$ for the weak feedback\footnote{Note that this transition is not
  relevant for our two limiting cases of H$_2$ cooling without
  feedback and of no H$_2$ cooling.}. Finally, at lower redshifts the
X-ray background heats the gas, with the transition at $1+z_h = 18.2$,
$16.9$, $16.3$,and $15.9$ for H$_2$ cooling without feedback, weak and
strong feedbacks and no H$_2$ cooling, respectively.

Naturally, the delayed star formation affects our predictions for the
expected global spectrum of the 21-cm signal from high redshifts by
blue-shifting the absorption trough.  As we see from
Figure~\ref{fig:T21}, the general shape of the global spectrum varies
only mildly as we change the strength of the feedback (or turn off
$\vbc$). For the most realistic cases (i.e., weak or strong feedback,
including $\vbc$), the spectrum reaches its minimum temperature of
$T_{21} = -121\pm 2$~mK at $\nu_{\textrm{min}} = 63\pm2$~MHz (compare
a shift of $\Delta\nu \sim -2.5$~MHz if $\vbc$ is turned off, and
$\Delta\nu \sim 6$~MHz if stars cannot form via $H_2$ cooling; see
Table~\ref{Tab:Global}).

The foreground galactic emission, which is the main noise component
after the terrestrial interference is removed, is expected to be
fairly smooth. Therefore it is easier to detect the signal if it
changes fast with frequency.  There are two characteristic frequencies
for which the global 21-cm spectrum changes fast: the strongest change
is at the low-frequency-end of the spectrum (high redshifts), at $\nu
_{\textrm{max,1}} = 54.5 \pm 1.5$~MHz (for the most realistic cases)
where the slope reaches $|dT_{21}/d\nu| = 9.6 \pm 1.1$~mK/MHz; whereas
the second strongest change happens at higher frequencies (low
redshifts), at $\nu_{\textrm{max,2}} = 72.8 \pm 1.9$~MHz, where the
slope is $|dT_{21}/d\nu| \sim 6.9$ mK/MHz. Clearly, while the shape of
the spectrum does not vary much with feedback (at least at fixed X-ray
and Ly$\alpha$ efficiencies, as in this work), the frequencies
$\nu_{\textrm{min}}$, $\nu_{\textrm{max,1}}$, and
$\nu_{\textrm{max,2}}$ substantially depend on the feedback strength
(and $\vbc$). Thus, measuring them may help us constrain the actual
strength of the LW feedback in Nature.

\section{Evolution of Fluctuations}

The power spectrum of $T_{21}$ may be easier to observe on top of the
smooth foregrounds than the global spectrum, and it contains more
information than the single curve. As we demonstrate further, the
shape of the 21-cm power spectrum evolves with redshift in a
non-trivial way and has the power to constrain high-z astrophysical
processes.

\begin{figure*}
\includegraphics[width=3.4in]{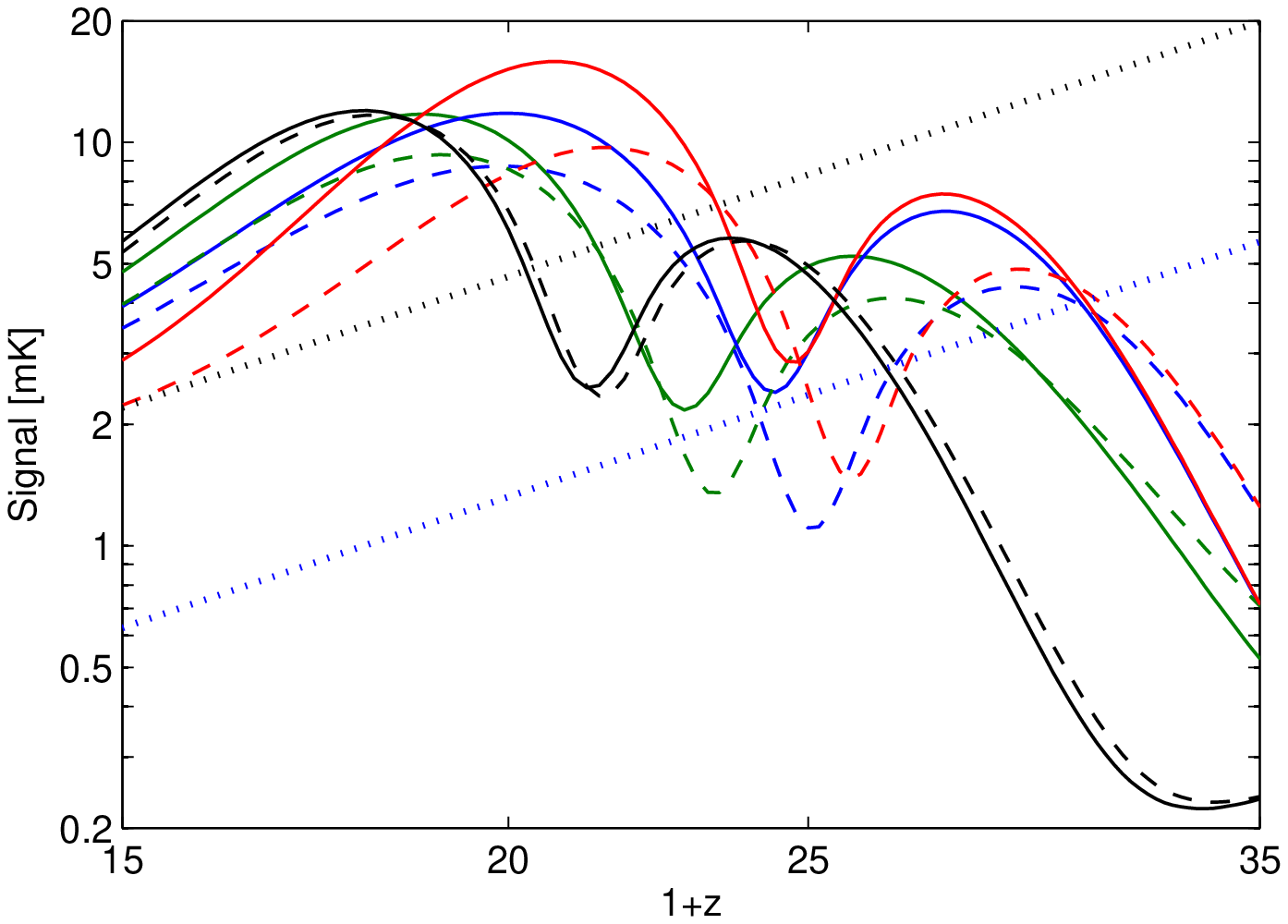}
\includegraphics[width=3.4in]{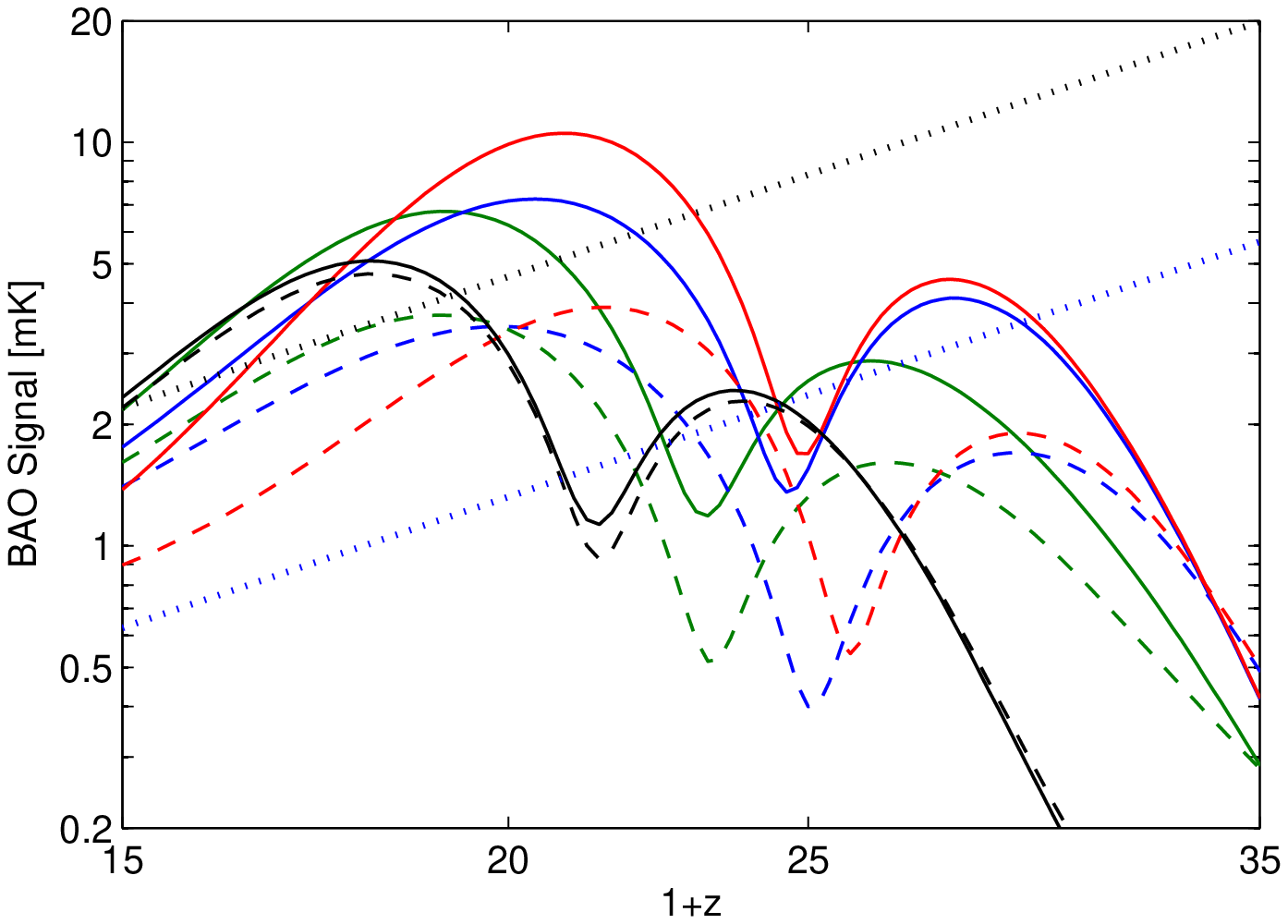}
\caption{{Left:} amplitude of the 21-cm brightness temperature at 125
  Mpc, the BAO scale. {Right:} peak-to-trough amplitude in BAOs, which
  we define to be the height of the BAO peak at $2 \pi/k=125$ Mpc
  minus the depth of the trough at $2 \pi/k=90$ Mpc (each feature
  measured with respect to a smoothed power spectrum with the BAOs
  removed). The two panels show the signal with (solid) and without
  (dashed) $\vbc$ for no H$_2$ cooling (black), strong (green) and
  weak (blue) feedback and for no feedback (red). We also show the
  sensitivity curves for SKA (blue dotted) and MWA/LOFAR-like (black
  dotted) experiments; the latter refers to an instrument with the
  same collecting area as MWA or LOFAR, and the former to the planned
  SKA, where to both we have applied an estimated degradation factor
  due to foreground removal (see \citet{McQuinn:2006} and
  \citet{Visbal:2012} for details).}
\label{fig:Sig1}
\end{figure*}

Interestingly, we find that the heating fluctuations in the 21-cm
signal are mostly decoupled from the Ly$\alpha$ fluctuations
\footnote{\citet{Pritchard:2007} argued that Ly$\alpha$ photons from
  X-ray ionization are significant, but they appear to have
  overestimated the X-ray intensity relative to Ly$\alpha$ due to
  normalization errors. We find that $\sim 100$ times higher X-ray
  intensity is needed to generate a significant Ly$\alpha$
  contribution from X-rays. }. To illustrate this phenomenon we show
in Figure~\ref{fig:Sig1} the redshift evolution of the 21-cm signal at
the BAO scale (125 Mpc) and the peak-to-trough signal of BAOs in the
brightness temperature. As in the case of the global signal, the shape
of the curve is determined by the interplay between X-rays and
Ly$\alpha$, while the main role of the LW feedback (and $\vbc$) is to
delay the evolution and enhance large-scale fluctuations (in
qualitative agreement with \citet{Visbal:2012} and
\citet{Fialkov:2013}). The signal has two clear peaks: one at high
redshifts (e.g., $z\sim 25$ in the case of strong feedback) which is
sourced by fluctuations in the Ly$\alpha$ background, and the other at
lower redshifts due to the fluctuations in the temperature of the gas
($z\sim 18$ for the strong
feedback).  
The exact values of the extremes of the signal and their corresponding
redshifts for all of the discussed cases are listed in
Table~\ref{Tab:Sig}. The clear separation between these two domains
means that the two regimes are disentangled, and that the onset of
Ly$\alpha$ fluctuations and of the heating era can be measured
separately. This would be a very powerful tool to explore
astrophysical processes at high redshifts, such as the emissivity of
the first galaxies in UV and X-rays, the halo masses that contribute
to star formation at each redshift and the role of the negative
feedback. Naturally, the spacing between the peaks would change if we
were to alter our model parameters.

\begin{table*}
\begin{center}
\begin{tabular}{ |l  | r |r| r|r| r| r| r|r|r|r|r|r|}
\hline
Feedback,  $\vbc$ &  S$_{\rm{max,2}}$ [mK] & $z$ & S$_{\rm{max,1}}$ [mK]& $z$ & S$_{\textrm{BAO},\rm{max,2}}$ [mK] & $z$  & S$_{\textrm{BAO},\rm{max,1}}$ [mK]& $z$ \\
\hline
None, no $\vbc$ & 9.7 & 19.6 &  4.9 & 27.4 & 3.9 & 19.6& 1.9 & 27.4 & \\
None & 15.8 & 18.8 &  7.5 & 25.6& 10.5 & 18.8 & 4.6 & 25.8 & \\
Weak, no $\vbc$ &8.9 & 18.0 & 4.4 & 27.2 & 3.5 & 17.9 & 1.7 & 27.2\\
Weak & 12.1 & 18.1 &  6.7& 25.8& 7.2& 18.4 & 4.1 & 25.8 & \\
Strong, no $\vbc$ & 9.5   & 17.0 & 4.1 & 24.6 & 3.7 & 17.0 & 1.6 & 24.6 \\
Strong & 12.0 & 16.8 &  5.2 & 23.8& 6.9 & 17.1 & 2.9 & 24.2 & \\
No H$_2$, no $\vbc$& 12.0 & 16.2 & 5.7 & 21.8 & 4.9 & 16.2 & 2.3 & 21.8 \\
No H$_2$ & 12.4 & 16.0&  5.8 & 21.6& 5.3 & 16.1 & 2.4 & 21.6& \\
\hline
\end{tabular}
\caption{\label{Tab:Sig} The maximal values of the signal in $T_{21}$
  and in BAO together with the corresponding redshifts during the
  X-ray heating era (low-z) and the Ly$\alpha$ coupling era (high-z).
  The Table summarizes the results from Figure~\ref{fig:Sig1}.}
\end{center}
\end{table*}

\begin{figure*}
\includegraphics[width=3.4in]{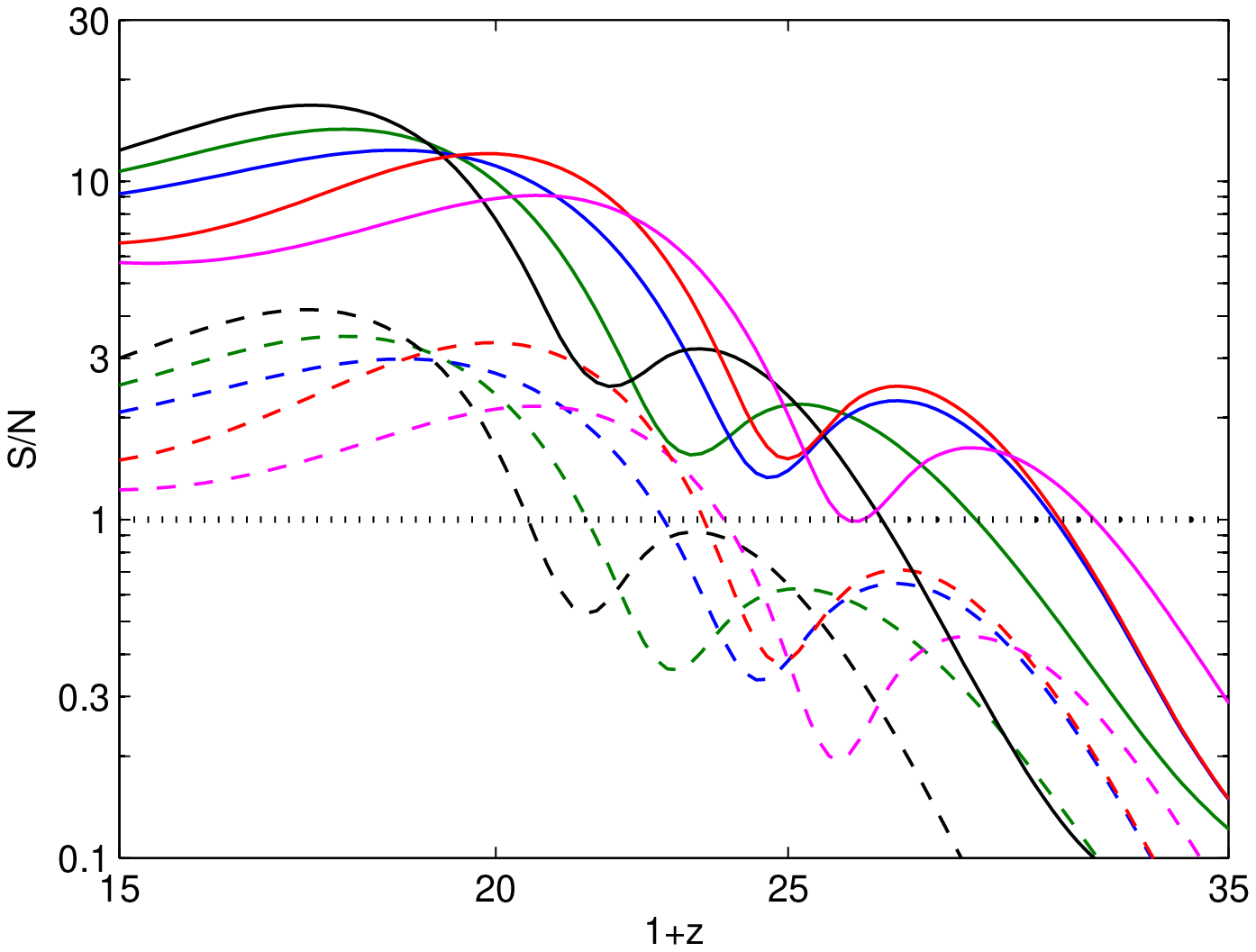}
\includegraphics[width=3.4in]{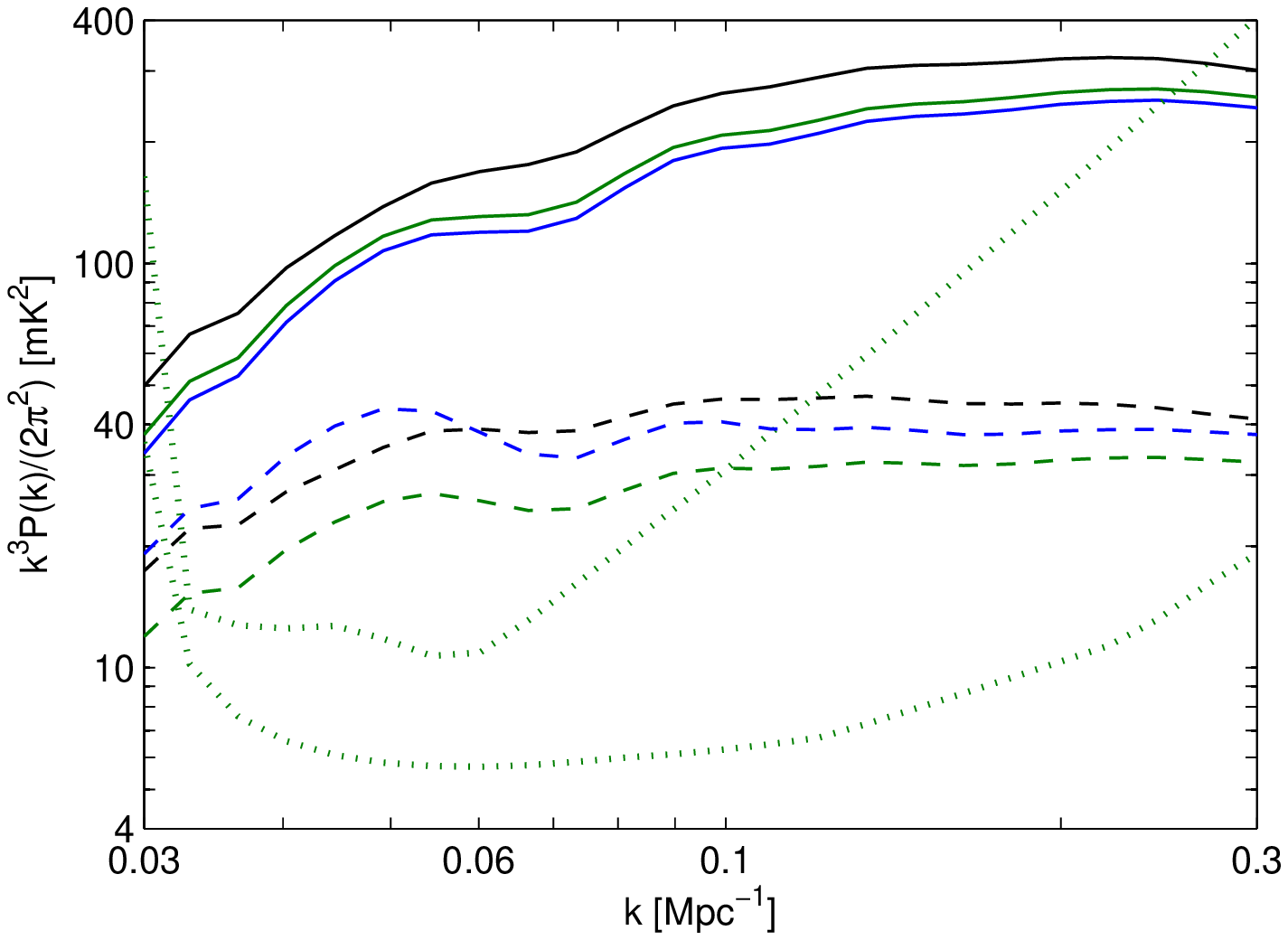}
\caption{{\bf Left:} Maximal S/N of the redshifted 21-cm signal at a
  single wavenumber $k$ in the range $k = 0.03-0.3$ Mpc$^{-1}$, where
  the signal is normalized by the noise of the SKA (solid lines) or by
  the noise of an MWA/LOFAR-like experiment (dashed lines).  The
  colors refer to no feedback (red), weak (blue) and strong (green) LW
  feedbacks or no H$_2$ cooling (black), and all cases include $\vbc$.
  We also show the previously standard case for small halos, which
  neglected both $\vbc$ and feedback (magenta curves). {\bf Right:} Power
  spectrum for the realistic cases (with $\vbc$) of strong (green) and
  weak (blue) feedbacks and no H$_2$ cooling (black) calculated at the
  corresponding redshifts of the low-z (solid) and the high-z (dashed)
  peaks in the S/N from the left panel of this Figure. The redshifts
  are $17.6$ and $26.2$ for the weak feedback, $16.8$ and $24.2$ for
  the strong feedback, and $16.4$ and $22.4$ for the no H$_2$ case.
  The upper dotted line correspond to the sensitivity curve of
  MWA/LoFAR at $z = 16.8$ (low-z S/N peak of the strong feedback
  case), while the lower dotted line corresponds to the sensitivity of
  SKA at $z = 24.2$, the redshift of the high-z S/N peak in the strong
  feedback case. Analogous sensitivity curves for the weak feedback
  case would be higher by a factor of about 1.3 (MWA/LoFAR) or 1.5
  (SKA); for no H$_2$ cooling case the curves would be lower by 1.1
  (MWA/LoFAR) or 1.5 (SKA). }
\label{fig:Sig2}
\end{figure*}

As was recently shown, heating fluctuations in the 21-cm power
spectrum from high redshifts ($z\sim 20$) are expected to be strong
enough to be detectable even with present-day technology
\citep{Visbal:2012}. Here we demonstrate (Figures~\ref{fig:Sig1} and
\ref{fig:Sig2} and Table~\ref{Tab:SN}) that the fluctuations seeded by
the inhomogeneous Ly$\alpha$ background are probably a bit too weak to
be observed with an MWA/LOFAR-like telescope, but are expected to be
strong enough to be detected by the next-generation SKA experiment. In
other words, SKA should have enough sensitivity to detect the complete
early history of the 21-cm fluctuations, for any scenario of the LW
feedback.

\begin{table}
\begin{center}
\begin{tabular}{ |l  | c|r| c|r| c| r| r|r|r|r|r|r|}
\hline
Feedback, $\vbc$ & S/N$^{\rm{MWA/LOFAR}}_{\rm{max,1}}$ & $z_{\rm{1}}$ & S/N$^{\rm{SKA}}_{\textrm{max,2}}$ & $z_{\rm{2}}$ \\
\hline
None, no $\vbc$ &    2.2 & 19.6 & 1.7 &   27.6\\
None &    3.4 & 19.0 & 2.6 &  26.2\\
Weak, no $\vbc$ &  2.5 & 17.6  & 1.6 &  27.5\\
Weak &  3.0 & 17.7 & 2.3 &   26.2 \\
Strong, no $\vbc$ &  3.0 &  17.1 & 1.9 &   25.0\\
Strong &    3.5 & 16.8 & 2.3  &   24.2 \\
No H$_2$, no $\vbc$&   4.1 &  16.5 & 3.2  &  22.6\\
No H$_2$ &   4.3 & 16.4 & 3.3   & 22.2\\
\hline
\end{tabular}
\caption{\label{Tab:SN}Maximal values of the signal to noise from
  Figure~\ref{fig:Sig2}. S/N$^{\rm{MWA/LOFAR}}_{\rm{max,1}}$ is the
  maximal S/N for an experiment like MWA or LOFAR for detection of
  heating fluctuations (low-z), while
  S/N$^{\rm{SKA}}_{\textrm{max,2}}$ is the maximal signal to noise for
  detection of fluctuations from Ly$\alpha$ coupling by the SKA.}
\end{center}
\end{table}

The shape of the power spectrum (Figure~\ref{fig:Sig2} right panel)
depends on the nature of the fluctuations (either from the heating era
or from the era of Ly$\alpha$ coupling) and on the type of the
feedback. The effect of the LW feedback and of $\vbc$ on heating
fluctuations (solid lines, right panel of Figure~\ref{fig:Sig2}) was
discussed in detail in \citet{Fialkov:2013} and \citet{Visbal:2012}.
Here we complete the picture by including the effect of the Ly$\alpha$
coupling (dashed lines, right panel of Figure~\ref{fig:Sig2}). In
fact, the strength of the feedback determines whether or not the LW
transition happens during the heating era or the epoch of Ly$\alpha$
coupling, which decides the character of the large-scale fluctuations.
For instance, if the feedback is weak, it becomes relevant only at low
redshifts, and affects mainly the X-ray heating, while only having a
minor effect on the Ly$\alpha$ fluctuations. On the other hand, if the
feedback is strong, then the LW radiation begins early on to suppress
the low-mass halos that are sensitive to the effect of $\vbc$, so the
BAOs are substantially suppressed even during the Ly$\alpha$
fluctuations. In any case, though, the LW feedback changes quite
gradually with redshift \citet{Fialkov:2013}.

Clearly, the nature of the source of the fluctuations plays a
significant role in shaping the power spectrum. In particular, the
power spectrum from the era of Ly$\alpha$ coupling is flat on scales
smaller than $\sim 100$ Mpc, which is of order the effective horizon
for the Ly$\alpha$ photons, within which the fluctuations are expected
to be suppressed. In the X-ray case, a larger fraction of the flux is
absorbed close to the source, and so the power spectrum is much
steeper. We note that recently \citet{Mesinger:2013} studied a model
similar to our no $H_2$, no $\vbc$ case; in this extreme case of no
star formation from $H_2$ cooling, we find that adding $\vbc$ still
shifts the key 21-cm features by up to a few percent.

To summarize, we have shown that the prospects for observing the
complete early history of the Universe via fluctuations in the 21-cm
signal are very promising. Future 21-cm observations are expected to
constrain the formation history and properties of stars and galaxies
at high redshifts. The effect of fluctuations in the Ly$\alpha$ and
X-ray backgrounds on the 21-cm power spectrum are likely to be mostly
decoupled in time and can be studied separately.  Also, the LW
feedback plays a unique role by delaying star formation and thus can
be constrained independently.

\section{Acknowledgments}
This work was supported by Israel Science Foundation grant 823/09.
A.F.\ was also supported by European Research Council grant 203247.


\end{document}